\documentclass[aps,prl,twocolumn,groupedaddress]{revtex4}

\usepackage{graphicx,psfrag}
\begin{document}













{\bf Reply to the comment of Jeng and Schwarz}.
In \cite{TBF} we introduced a class of kinetically constrained 
models which display a dynamical glass transition:  above a critical density, $\rho_c$, 
there appears an infinite
cluster of {\it frozen} particles which can never  be moved.   At $\rho_c$ 
the density of frozen particles, $\phi(\rho)$, is discontinuous, while as $\rho\nearrow\rho_c$ 
 there is an exponentially diverging crossover length, $\Xi(\rho)$.
This   {\it jamming percolation} behavior in two dimensions is a consequence of two perpendicular  directed-percolation (DP)-like processes which together can form a frozen network of DP segments ending at T junctions with perpendicular DP segments.

In \cite{TBF}, we focused on a particular example: the ``knights" model.  As correctly pointed out  by Jeng and Schwarz (JS) \cite{comment},  we overlooked
some additional directed frozen structures of the knights model which are {\it not} 
simple
DP  paths: these ``thicker" directed structures lower the critical density. Here we argue that, nevertheless, the full directed processes are in the DP universality class and the T-junctions between perpendicular segments of these give rise to a jamming percolation transition with the 
universal properties discussed in \cite{TBF}. 
Moreover, although for the knights our results are not rigorous, for some other models they {\sl are} rigorous.

The simplest is the ``spiral" 
model. It is similar to the knights model except that blocking of a particle is by either its N, NE and S, SW, or its W, NW and E, SE pairs of  
neighbors as in Fig. (b) ({\it c.f.} Fig. 1(b) of \cite{TBF}) .
As for the knights model, there are two directed processes: one in the NNE-SSW
 and the other in the ESE-WNW direction. But now these are exactly congruent \cite{TB} to site DP processes on a square lattice. As infinite occupied DP paths are {\it necessary and sufficient} for an infinite frozen structure,
the critical density for the spiral
model is $\rho_c^{DP}$, 
and all our results \cite{TBF} 
are rigorous (up to standard conjectures on the anisotropic scaling of DP)  \cite{TB}.
   
We claim that the knights model is in the same universality class as the 
spiral model. 
Analysis of the blocking rules yields, for each diagonal
 direction, 
two infinite sequences of  thicker and thicker DP  
processes, {\sl SDP} and {\sl NDP} for which the existence of percolating occupied paths 
are, respectively {\it sufficient} and  {\it necessary} for an infinite blocked cluster of the knights model. JS's structures are in the {\sl SDP} sequence.
We conjecture that the limits of {\it both} sequences belong to the DP universality class and the limit points
of their critical densities coincide at some $\rho_c^\infty$. Thus for the knights model there are two perpendicular sets of  DP paths and effective T-junctions between them causing  
 a jamming transition at $\rho_c^{\rm knights}=\rho_c^\infty$.
To test this 
we performed simulations on two types of long diagonal strips of length
$L$ and width $W\propto L^\zeta$, with $\zeta\simeq 0.63$ the DP anisotropy exponent.  
Boundary conditions empty on the sides and filled on top and bottom
focus on {\sl SDP} paths:   $P_S(\rho,L)$ is the probability of a frozen spanning cluster.
Boundary conditions filled on  one side but empty on the other side and on top 
and bottom focus on {\sl NDP} paths (which are those preventing the arbitrary expansion of large holes \cite{TB}): $P_N(\rho,L)$ is the probability of some frozen particles in the {\it open half} of such strips.
Both the $P_{\sl S}$ and $P_{\sl N}$ data cross at the {\it same} 
critical density, $\rho_c\simeq 0.6359$, and display good scaling with
$(\rho-\rho_c)L^{1/\nu}$, where $\nu\simeq 1.73$ is the parallel correlation length exponent for DP: Fig.(a) shows data for $P_{\sl S}$. We find the {\it same} behavior for the spiral model (with a different $\rho_c$).
This
  yields strong support for the conjectured universality of the DP-like processes on large length scales. Thus the arguments in \cite{TBF} for the dynamical glass transition can be applied.
As predicted,
the density of {\sl frozen} particles, $\phi$, is non-zero at $\rho_c$:
 Fig. (c) shows $\phi(\rho_c,L)$ for knights model. This shows that the T-junction interactions between the two DP processes are crucial as the density of DP clusters
vanishes at the DP transition.\\
To summarize 
our results in \cite{TBF} are rigorous for the spiral model with $\rho_c=\rho_c^{DP}$.
For the knights models $\rho_c\neq \rho_c^{DP}$ since JS's and more complicated directed processes are not simple DP processes \cite{comment}. Nevertheless, our numerical results  strongly suggest that the critical behavior remains the same.

C. Toninelli$^{*}$, G. Biroli$^{\dagger}$, D.S. Fisher$^{**}$.
* LPMA,
CNRS-UMR 7599, Univ.Paris VI-VII, 4 Pl.Jussieu, Paris, FRANCE, 
$^{\dagger}$ SPhT, CEA/Saclay-Orme des Merisiers, F-91191 Gif-sur-Yvette Cedex, FRANCE. $^{**}$ Lyman Laboratory of Physics, Harvard University, Cambridge, MA 02138, USA.

\begin{figure}[h]
\psfrag{rhoc}[][][4]{$\phi(\rho_c)$}
\psfrag{x}[][][2]{x}
\psfrag{PBT}[][][3]{{{$P_{\sl S}$}}}
\psfrag{P}[][]{{\tiny{SE}}}
\includegraphics[angle=270, width=.9\columnwidth]{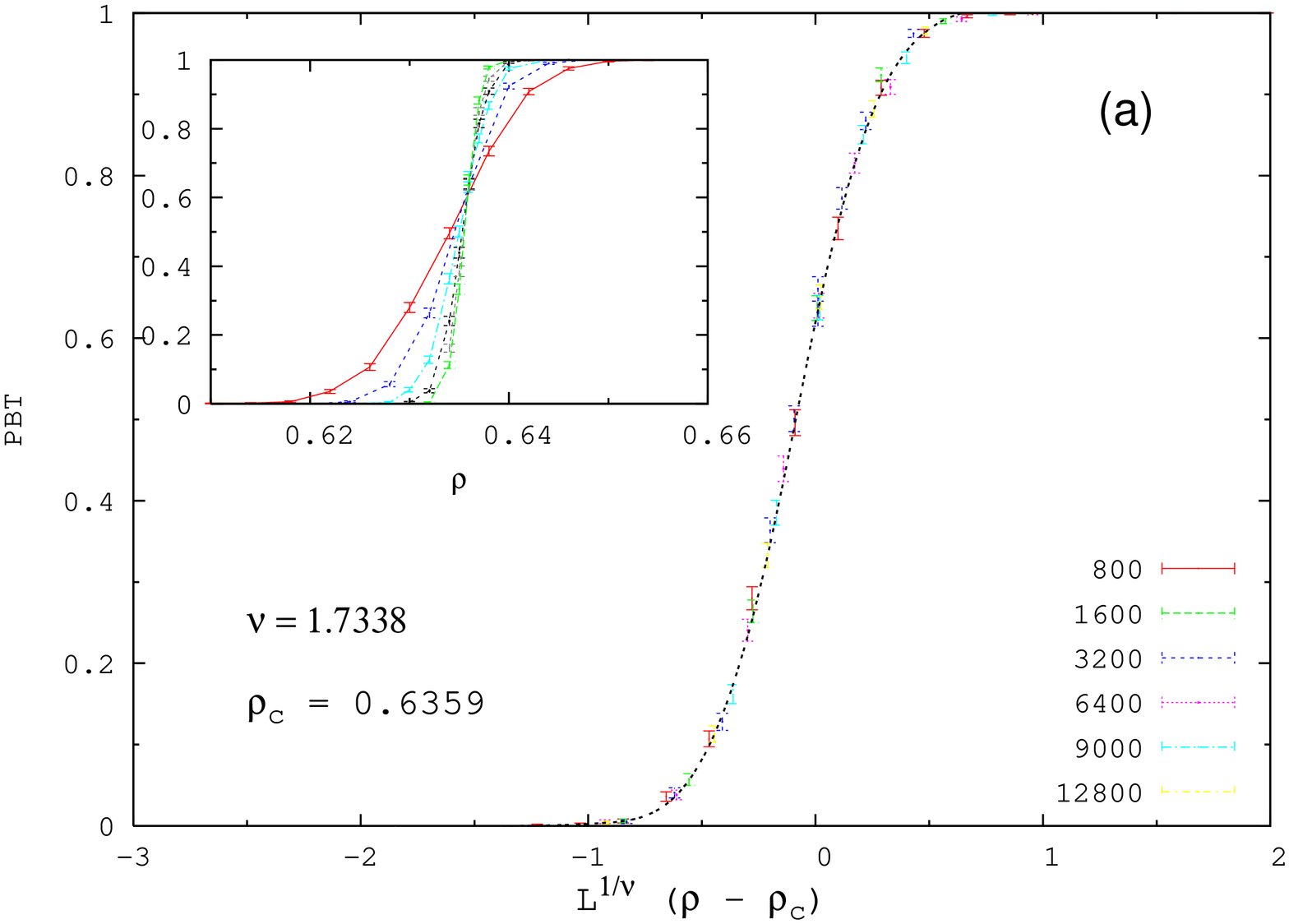}
\includegraphics[angle=270,width=0.3\columnwidth]{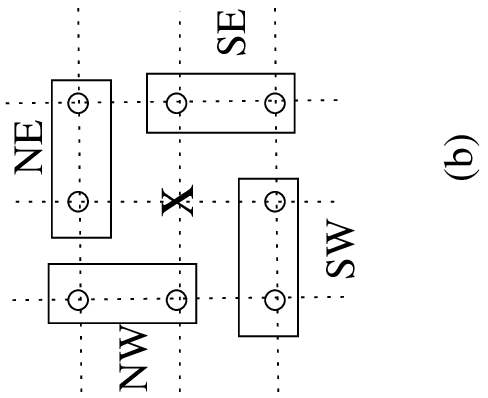}
\includegraphics[angle=270,width=0.6\columnwidth]{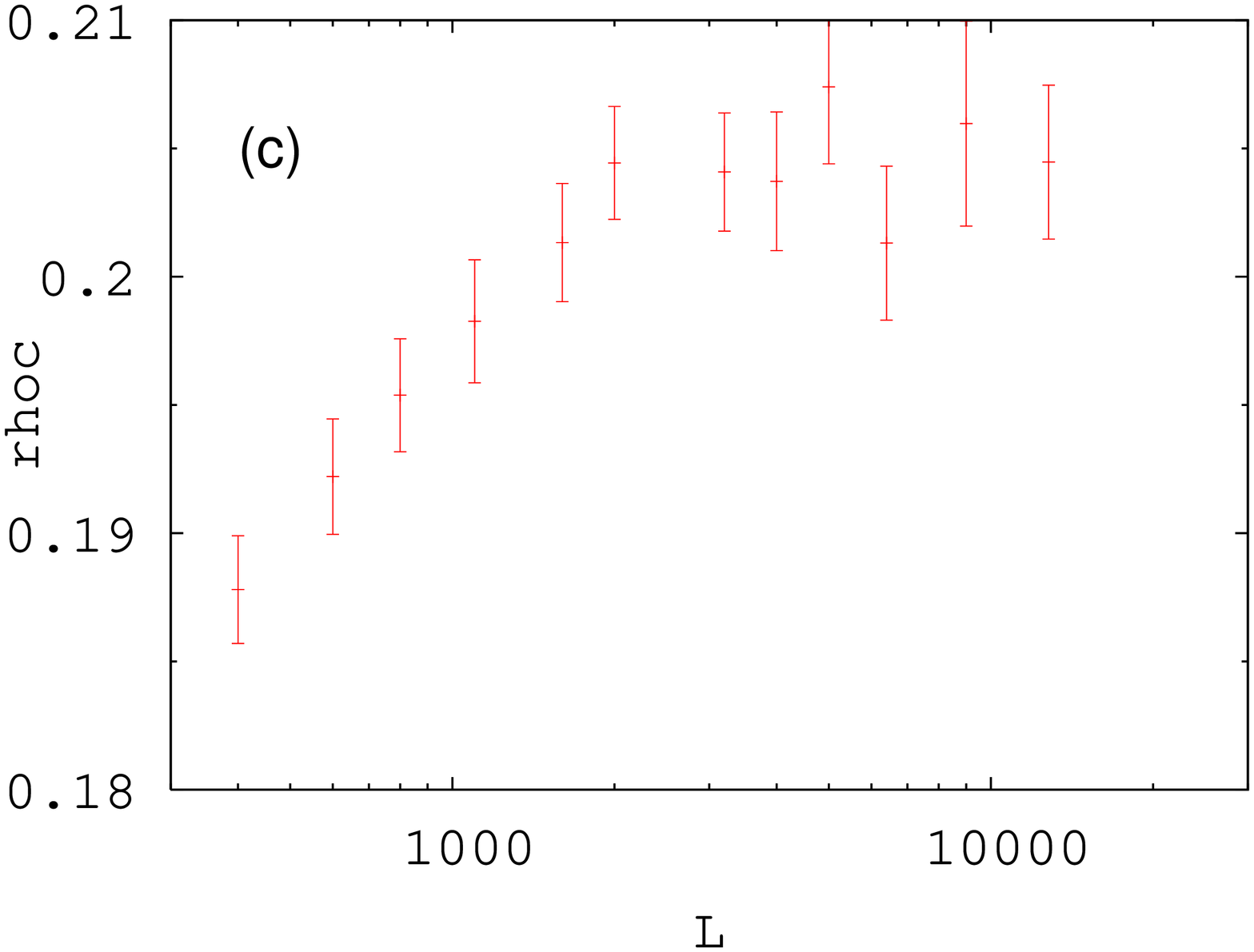}
\end{figure}

\end{document}